# Can Neutron Star Tidal Effects Obscure Deviations from General Relativity?

Stephanie M. Brown[1,2,3], Badri Krishnan[1,2,4], Rahul Somasundaram[5,6], and Ingo Tews[5]
[1] Albert-Einstein-Institut, Max-Planck-Institut für Gravitationsphysik, Callinstraße 38, 30167 Hannover, Germany
[2] Leibniz Universität Hannover, 30167 Hannover, Germany
[3] Oskar Klein Centre, Department of Physics, Stockholm University, AlbaNova University Centre, SE 106 91 Stockholm, Sweden
[4] Institute for Mathematics, Astrophysics and Particle Physics, Radboud University Heyendaalseweg 135, 6525 AJ Nijmegen, The Netherlands
[5] Theoretical Division, Los Alamos National Laboratory, Los Alamos, NM 87545, USA
[6] Department of Physics, Syracuse University, Syracuse, NY 13244, USA

## Abstract

One of the main goals of gravitational-wave astrophysics is to study gravity in the strong-field regime and constrain deviations from general relativity (GR). Any such deviation affects not only binary dynamics and gravitational-wave emission but also the structure and tidal properties of compact objects. In the case of neutron stars, masses, radii, and tidal deformabilities can all differ significantly between different theories of gravity. Currently, the measurement uncertainties in neutron star radii and tidal deformabilities are quite large. However, much less is known about how the large uncertainty in the nuclear equation of state (EOS) might affect tests of GR using binary neutron star mergers. Conversely, using the wrong theory of gravity might lead to incorrect constraints on the nuclear EOS. Here, we study this problem within scalar–tensor (ST) theory. We apply the recently derived $\ell = 2$ tidal Love numbers in this theory to parameter estimation of GW170817. Correspondingly, we test if physics beyond GR could bias measurements of the nuclear EOS and neutron star radii. We find that parameter inference for both the GR and ST cases returns consistent component masses and tidal deformabilities. The radius and the EOS posteriors, however, differ between the two theories, but neither is excluded by current observational limits. This indicates that measurements of the nuclear EOS may be biased and that deviations from GR could go undetected when analyzing current binary neutron star mergers.

*Unified Astronomy Thesaurus concepts:* Gravitational waves (678); Gravitational wave astronomy (675); Neutron stars (1108); Gravitation (661); Non-standard theories of gravity (1118)

## 1. Introduction

Gravitational waves (GWs) from a compact binary merger were first detected in 2015 by the LIGO and Virgo observatories (B. P. Abbott et al. 2019a). Since the first detection, $\mathcal{O}(10^2)$ compact binary mergers have been observed. Most of these detections are binary black hole systems, but there are a handful of binary neutron star and neutron star–black hole mergers as well. A merger of two neutron stars was first observed in 2017, and the event, referred to as GW170817, was observed simultaneously in the electromagnetic spectrum (B. P. Abbott et al. 2017a, 2017b, 2017c). These GW events have been used in various tests of general relativity (GR). See, e.g., R. Abbott et al. (2021a), B. P. Abbott et al. (2016, 2019a, 2019b), A. K. Mehta et al. (2022), Y.-F. Wang et al. (2021), K. Chatziioannou et al. (2021), R. Nair et al. (2019), S. Mirshekari & C. M. Will (2013), C. D. Capano et al. (2023), and M. Isi et al. (2019, 2021). These tests probe various aspects of the inspiral, merger, and ringdown regimes of compact binary coalescence. These include tests of post-Newtonian theory in the inspiral regime, consistency tests across the merger, and properties of quasi-normal modes in the post-merger regime. The observation of binary neutron star systems, especially GW170817, made it possible to study neutron stars and their properties with GWs.

Many of these observational tests are "nonparametric," in the sense that they probe the consistency of predictions by standard GR with observational data without the need for a well-developed alternate theory of gravity. These include tests of the black hole no-hair theorems (M. Isi et al. 2019; C. D. Capano et al. 2023), the area increase law (M. Isi et al. 2021; S. Kastha et al. 2022), and the consistency between inspiral and merger (A. Ghosh et al. 2016). The so-called parametric tests carried out so far are based on the consistency of the post-Newtonian parameters; see, e.g., R. Abbott et al. (2021a) and B. P. Abbott et al. (2016, 2019a, 2019b). These tests, including parameterized post-Newtonian and parameterized post-Einsteinian theories, constrain deviations from the post-Newtonian signal (C. M. Will 1971; C. K. Mishra et al. 2010; A. Gupta et al. 2020; M. Saleem et al. 2022) but are currently not based on a specific well-developed fundamental alternative to standard GR. There are, of course, several alternatives available in the literature, but none of these have been studied to the same extent as GR, and none can be considered sufficiently compelling at the exclusion of the others. Several alternate theories have been ruled out entirely by comparing the propagation of the GW signal in GW170817 with the propagation of signals from the electromagnetic counterparts (B. P. Abbott et al. 2017c, 2019a).

Here, we are concerned with strong-field effects that modify the structure of the compact objects. Even if there were a compelling and well-developed alternative to GR with accurate gravitational waveforms available, challenging GR would remain a daunting task. The high degree of confidence required for such a claim would require that numerous systematic errors in the data calibration, detector noise artifacts, waveform modeling, etc., are all thoroughly addressed and understood.

In this work, we have a less ambitious, but nevertheless important, goal: to understand how ignoring deviations from GR can lead to biases in interpreting gravitational-wave events. Specifically, we consider constraints on the nuclear equation of state (EOS) and neutron star radii inferred from GW170817. These constraints are based on the tidal deformations of neutron stars under an external perturbation. If GR is not the correct theory for gravity, then our estimates of neutron star radii and tidal deformabilities might be incorrect. The concrete questions addressed here are: how large could such biases be? Can the nuclear physics parameters used to construct the EOS be constrained, and which parameters are most likely to be constrained? How would these constraints be affected by considering an alternate theory of gravity? We address these questions in the context of scalar–tensor (ST) theories of gravity. By measuring neutron star radii with a particular ST theory, using GW data from GW170817, we show that these measurements can indeed be biased by ignoring deviations from GR.

## 2. Neutron Star Tidal Love Numbers

During the inspiral phase of a binary neutron star merger, both neutron stars are tidally deformed, and the emitted gravitational-wave signal carries an imprint of this deformation. The tidal deformability and the associated tidal Love numbers (TLNs) characterize the response or deformation of the neutron star to an external tidal field, such as the one generated by a binary companion (E. E. Flanagan & T. Hinderer 2008; T. Hinderer 2008; T. Binnington & E. Poisson 2009; T. Damour & A. Nagar 2009). Consider a neutron star of mass $M$ and radius $R$. The tidal deformability $\lambda_\ell$ relates the induced $\ell$-multipole moment and the external tidal field, and it depends on the composition of the star. The dimensionless tidal deformabilities are in turn related to the dimensionless TLN $k_\ell$:

$$\lambda_\ell = \frac{2}{(2\ell - 1)!!} k_\ell \left(\frac{R}{M}\right)^{2\ell+1}. \quad (1)$$

For a black hole, the tidal deformability is generally believed to vanish, but for a neutron star it depends on the neutron star EOS. At leading order, the dominant effect is quadrupolar ($\ell = 2$), and the tidal effects on the GW signal from a binary system with component masses $M_{1,2}$ are captured by the combination

$$\tilde{\lambda} = \frac{16}{13} \frac{(12q + 1)\lambda_2^{(1)} + (12 + q)q^4 \lambda_2^{(2)}}{(1 + q)^5}, \quad (2)$$

where $\lambda_\ell^{(1,2)}$ are the tidal deformability parameters for each of the stars, and $q = M_2/M_1 \leqslant 1$ is the mass ratio. An estimation of $\tilde{\lambda}$ leads to constraints on the EOS, and thus, on neutron star radii (B. P. Abbott et al. 2018; S. De et al. 2018; C. D. Capano et al. 2020). Analyses of GW170817 have provided improved constraints on the tidal deformability and therefore on the nuclear EOS, which determines the neutron star's structure and therefore its observable properties such as its mass, radius, and tidal deformability (B. Abbott et al. 2018; D. Radice & L. Dai 2019; G. Raaijmakers et al. 2021; P. T. Pang et al. 2023; M. Breschi et al. 2024). Despite recent improvements, there are large uncertainties in the nuclear EOS due to large uncertainties in the present data and the difficulty of robustly calculating the EOS starting from a first principles description provided by quantum chromodynamics.

The nuclear EOS is related to the mass, radius, and tidal deformability of the neutron star by a system of equations comprising the Tolman–Oppenheimer–Volkoff (TOV) equations and a second-order differential equation for the tidal deformability derived within GR. However, the structure equations differ between various theories of gravity. Hence, it is not surprising that recent work has found that neutron star properties also differ between theories of gravity (S. S. Yazadjiev et al. 2018; A. Saffer & K. Yagi 2021; S. M. Brown 2023; G. Creci et al. 2023). Nevertheless, these deviations from GR may be obscured by the large uncertainty in the nuclear EOS. This means that careful consideration of non-GR tidal effects is necessary for precision tests of GR.

## 3. Neutron Stars in Scalar–Tensor Theory

ST theories are among the most straightforward and best-studied alternatives to GR. In ST theories with a single scalar field, the metric tensor ($g_{\mu\nu}$) is coupled to a scalar field ($\varphi$) (P. Jordan 1955; M. Fierz 1956; C. Brans & R. H. Dicke 1961; Y. Fujii & K.-I. Maeda 2003). For a certain subset of these theories, nonperturbative effects are also shown to arise (T. Damour & G. Esposito-Farèse 1992, 1993). Such nonperturbative effects are invisible in standard post-Newtonian approximations but might be detectable through their effects on the tidal properties of neutron stars. ST theories can be formulated in either the Jordan or Einstein frame, with the metrics in the two frames related by a conformal rescaling. If $g_{\mu\nu}$ is the metric in the Einstein frame, then it is related to the Jordan frame metric $\tilde{g}_{\mu\nu}$ by $g_{\mu\nu} = A(\varphi)^2 \tilde{g}_{\mu\nu}$, where $A(\varphi)$ defines a specific theory. The metric $g_{\mu\nu}$ determines spacetime geometry, while $\tilde{g}_{\mu\nu}$ determines coupling with matter fields. The Einstein frame is generally used for computations because the equations simplify greatly in this frame. In the Einstein frame, the action takes the following form:

$$S = \frac{1}{16\pi G_*} \int \sqrt{-g} (R - 2g^{\mu\nu}\partial_\mu\varphi\partial_\nu\varphi - 2V(\varphi))d^4x + S_m[\Psi_m, A^2(\varphi)g_{\mu\nu}]. \quad (3)$$

Here, $\varphi$ is the Einstein frame scalar field, $R$ is the Ricci scalar, and $V(\varphi)$ is the scalar potential, which is set to 0 here. $S_m$ denotes the action of the matter, which is a function of the matter fields $\Psi_m$ and the Jordan metric $\tilde{g}_{\mu\nu}$.

The tidal deformability, which relates the induced multipole moment to the external perturbing field, depends on a second-order differential equation that defines the perturbations. In GR, there are two types of perturbations: even and odd tensor perturbations. The even parity or electric perturbations give rise to the familiar electric Love number $\lambda_\ell$. In ST theory, it is necessary to consider also the perturbations of the scalar field. This means there are two even-parity perturbations (scalar and tensor) and one odd parity perturbation. The tidal deformability associated with the scalar field $\lambda_\ell^S$ is dipolar at leading order (P. Pani & E. Berti 2014; S. M. Brown 2023; G. Creci et al. 2023). For this analysis, we focus on the even-parity tensor perturbation, which is analogous to its general relativistic counterpart.

Integrating the structure equations along with a nuclear EOS yields the mass–radius relationship, which is critical for astrophysical observations and gravitational-wave parameter estimation. By including the perturbation equation discussed in the previous paragraph, it is possible to calculate the tidal deformability as well. Each mass–radius–tidal deformability relationship is governed by the central density $\rho_c$, the scalar-field coupling constant, and the scalar field at infinity $\varphi_\infty$. As the boundary conditions are defined at the center of the star ($\rho_c$) and at infinity ($\varphi_\infty$), we use the shooting method to determine the correct value of the scalar field at the center of the star and then integrate the structure equations, the nuclear EOS, and the equation for tidal perturbations together. For details on the computation, see, e.g., S. M. Brown (2023). For all EOSs, we use the same scalar-field parameters in order to ensure that all mass–radius–tidal deformability relations used in the analysis refer to a consistent theory of gravity.

Our results are based on the spontaneous scalarization or Damour–Esposito-Farèse (DEF) model studied in T. Damour & G. Esposito-Farèse (1993). As mentioned above, this work demonstrated the existence of nonperturbative effects in neutron stars in ST theories. The DEF model is defined by the conformal factor

$$A(\varphi) = e^{\beta\varphi^2/2}, \quad (4)$$

with $\beta$ being a constant parameter. The parameters $\varphi_{\rm inf}$ and $\beta$ define a specific ST theory. When $\beta$ is sufficiently negative, even small values of $\varphi_\infty$ lead to large values of the scalar field in the neutron star. Thus, while the presence of the scalar field might be difficult to infer from the post-Newtonian regime, it might be revealed via finite-size effects of compact objects. Since the $\ell = 2$ electric tidal deformabilities are the most important finite-size effect for binary neutron star mergers, the DEF model provides a valuable testing ground for studying the impact of modified gravity theories. In this work, we shall take $\beta = -6$, which is sufficiently negative to cause nonperturbative effects in the interior of the star. The coupling strength, $\beta = -6$, is chosen because it represents the largest reasonable deviation from GR in the context of DEF. Any more negative couplings are ruled out, and any smaller couplings would lead to smaller effects than those discussed here. Scalarization only occurs for $\beta < -4.35$ (T. Harada 1998), effectively constraining the range of interest to $-6 \leqslant \beta \leqslant -4.35$. The value of the scalar field at infinity is constrained by the Cassini experiment (B. Bertotti et al. 2003), which constrains the $\varphi_\infty$ to $< 2.0 \times 10^{-3}$ when $\beta = -6$. We use $\varphi_\infty = 10^{-3}$ for this analysis. However, the TLNs are not strongly sensitive to this parameter (S. M. Brown 2023).

## 4. The Nuclear Equation of State

The construction of the EOS used here is similar to that of H. Koehn et al. (2024) and B. T. Reed et al. (2024), i.e., for the EOS below $2n_{\rm sat}$, where $n_{\rm sat}$ is the nuclear saturation density, we employ the metamodel of J. Margueron et al. (2018a, 2018b). The metamodel is a density-functional approach that allows one to directly incorporate nuclear physics knowledge encoded in terms of nuclear empirical parameters (NEPs). These parameters are defined via a Taylor expansion of the energy per particle in symmetric matter, $e_{\rm sat}$, and the symmetry energy, $e_{\rm sym}$, about saturation density:

$$e_{\rm sat}(n) = E_{\rm sat} + K_{\rm sat}\frac{x^2}{2} + Q_{\rm sat}\frac{x^3}{3!} + Z_{\rm sat}\frac{x^4}{4!} + \ldots, \quad (5)$$

$$e_{\rm sym}(n) = E_{\rm sym} + L_{\rm sym}x + K_{\rm sym}\frac{x^2}{2} + Q_{\rm sym}\frac{x^3}{3!} + Z_{\rm sym}\frac{x^4}{4!} + \ldots, \quad (6)$$

where $x \equiv \frac{n - n_{\rm sat}}{3n_{\rm sat}}$ is the expansion parameter. In this work, we fix the isoscalar NEPs describing symmetric matter to be $E_{\rm sat} = -16\,{\rm MeV}$, $n_{\rm sat} = 0.16\,{\rm fm}^{-3}$, and $Q_{\rm sat} = Z_{\rm sat} = 0\,{\rm MeV}$. Only the parameter $K_{\rm sat}$ is allowed to vary in the range 210–260 MeV. Regarding the isovector NEPs describing the symmetry energy, the lower-order parameters are varied in the ranges $E_{\rm sym} = [28, 35]$, $L_{\rm sym} = [30, 75]$, and $K_{\rm sym} = [-200, 200]\,{\rm MeV}$, whereas the parameters $Q_{\rm sym}$ and $Z_{\rm sym}$ are fixed at 0 MeV.

Above $2n_{\rm sat}$, we extend the neutron star EOS generated from the metamodel using the sound speed parameterization (I. Tews et al. 2018; R. Somasundaram et al. 2023). In this paper, we consider a fixed density grid at $[3, 4, 5, 6, 7, 8]n_{\rm sat}$, and the squared sound speed at each of these grid points is a free parameter of the model and is uniformly varied between 0 and $c^2$, where $c$ is the speed of light. Therefore, our EOS is determined by 10 parameters—four for the low-density EOS and six for the high-density EOS. By varying the parameters in their respective uniform ranges, we generate a set of 2500 EOSs, while ensuring that each EOS satisfies the constraint that the maximal neutron star mass $M_{\rm TOV} > 1.9\,M_\odot$.

## 5. Gravitational-wave Analysis

Bayesian inference is used to measure the posterior probability distribution $p(\boldsymbol{\theta}|h, \boldsymbol{d}(t); I)$ of the binary parameters $\boldsymbol{\theta}$. Here, $h(t)$ is the GW waveform and $\boldsymbol{d}(t)$ is the GW data from the LIGO-Hanford, LIGO-Livingston, and Virgo detectors. Additional information, such as the nuclear EOS, is denoted as $I$. The parameters $\boldsymbol{\theta}$ consist of six intrinsic parameters: the component masses $m_{1,2}$, the dimensionless-spin magnitudes $\chi_{1,2}$, and the tidal deformabilities $\lambda_{1,2}$ of each component star. In addition, it also contains seven extrinsic parameters, namely the sky location $(\alpha, \delta)$ with $\alpha$ being the R.A. and $\delta$ the decl., luminosity distance $d_L$, inclination angle $\iota$, coalescence time $t_c$, reference phase $\varphi$, and polarization angle $\psi$.

The methodology used here is the same as that developed in C. D. Capano et al. (2020). It uses the glitch subtracted time series data available from GW Open Science Center (R. Abbott et al. 2021b). The posterior probability distribution is calculated using Bayes' theorem $p(\boldsymbol{\theta}|\boldsymbol{d}(t), h) \propto p(\boldsymbol{d}(t)|\boldsymbol{\theta}, h)p(\boldsymbol{\theta})$, where $p(\boldsymbol{d}(t)|\boldsymbol{\theta}, h)$ is the likelihood of obtaining the data $\boldsymbol{d}(t)$ given a set of variables $\boldsymbol{\theta}$ and a waveform $h(t)$, and $p(\boldsymbol{\theta})$ is the prior distribution on the variables. We use the DYNESTY nested sampling algorithm (J. S. Speagle 2020) accessed through the PYCBC Inference framework (C. M. Biwer et al. 2019). Our sampling procedure is the same as in C. D. Capano et al. (2020). Using DYNESTY allows us to extract the evidence ($p(\boldsymbol{d}(t)|h)$) and calculate the Bayes factor, $\mathcal{B}$, which can be used to determine whether the data support one hypothesis over another. In this case, we define the Bayes factor to be the ratio of the

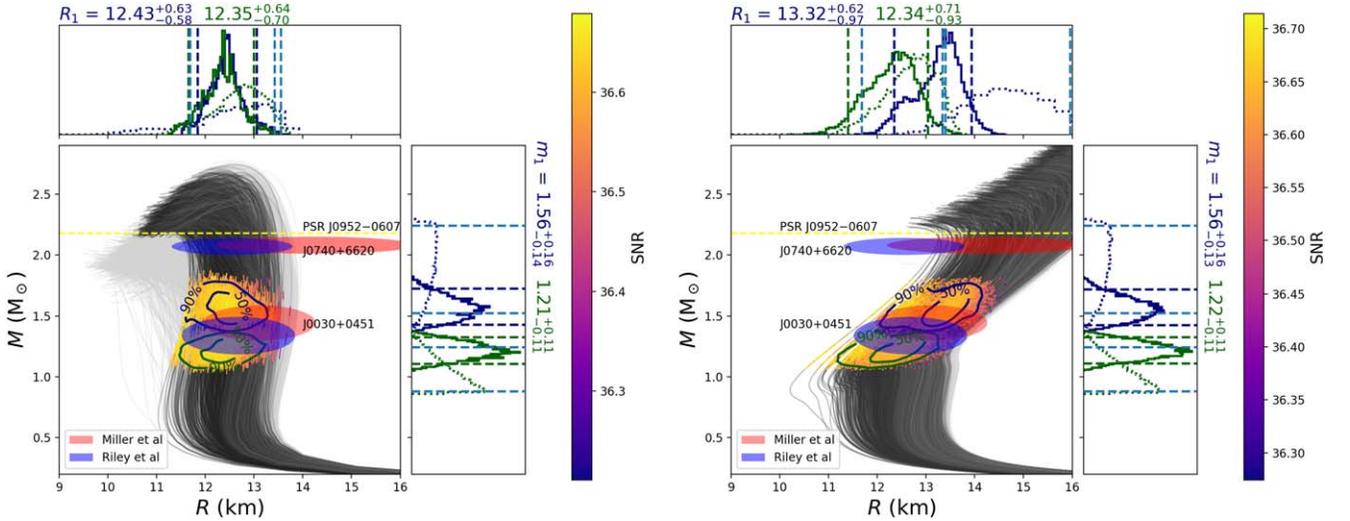

**Figure 1.** Neutron star mass–radius relations and marginalized posterior distributions of the source component masses $M_{1,2}$ and radii $R_{1,2}$, in GR (left) and ST theory with $\beta = -6$ (right). The shaded gray mass–radius curves show the prior distribution on the equation of state. The blue (green) curves on the mass–radius plot show the 50th and 90th percentile credible regions for the primary (secondary) mass. The mass and radius priors are shown in dotted lines in the 1D marginal plots. Additionally, electromagnetic constraints from the high mass pulsar PSR J0952−0607 (yellow dotted line) and NICER pulsars are J0740+6620 and J0030+0451 are shown (red and blue ovals). The ovals represent the $1\sigma$ bounds on the mass and radius and are not the posteriors from the NICER analysis. The color of the ovals denotes the source of the constraint.

evidences for ST theory over GR:

$$\mathcal{B} = \frac{p(\boldsymbol{d}(t)|h_{STT})}{p(\boldsymbol{d}(t)|h_{GR})}. \quad (7)$$

For the waveform model, we use IMRPhenomPv2_NRTidal (T. Dietrich et al. 2019b, 2019a). The mass priors are defined to be uniform in chirp mass ($\mathcal{M} = (m_1 m_2)^{3/5}/(m_1 + m_2)^{1/5}$) and symmetric mass ratio ($\eta = m_1 m_2/(m_1 + m_2)^2$). The spin priors assume a small, aligned spin and are uniform on $(-0.05, 0.05)$. The coalescence time is taken to be uniform from 0.1 s before to 0.1 s after the trigger time (118708882.42), $\cos\iota$ is taken to be uniform over $[-1, 1]$, and $\psi$ is taken to be uniform between $(0, 2\pi)$. We then randomly draw a specific EOS from the 2500 realizations that we have constructed and compute the tidal deformabilities using the component masses and the EOS, ensuring that both components have the same EOS. The sky location and luminosity distance $d_L$ are set to those of the observed electromagnetic counterpart: $\alpha = 13^h 09^m 48.1^s$, $\delta = -23° 22'53''.4$ (M. Soares-Santos et al. 2017) and $d_L = 40.7$ Mpc (M. Cantiello et al. 2018).

To compare the gravitational-wave parameter estimation results to current pulsar measurements, we include two pulsars measured by NICER: PSR J0030+0451 (M. C. Miller et al. 2019; T. E. Riley et al. 2019) and PSR 0740+6620 (M. C. Miller et al. 2021; T. E. Riley et al. 2021). We reject any EOS that is not within the $1\sigma$ bounds on the mass and radius for at least one of the pulsars. Additionally, we require that the EOS support the most massive pulsar observed to date (PSR J0952−0607). We do a strict cutoff at 2.18 $M_\odot$, which is the $1\sigma$ lower bound (R. W. Romani et al. 2022).

## 6. Results and Discussion

Our constraints on the radii and EOS are shown in Figure 1, where the left panel shows the results with standard GR and the right panel shows the results with ST gravity following the DEF model with $\beta = -6$. We find that the EOSs are indeed significantly modified within the particular ST theory that we have considered, especially for higher masses, which could lead to changes in the inferred radii.

We find that the component masses do not differ significantly, and it is reasonable to suppose that more moderate values of $\beta$ will also lead to similar masses. The radii, however, show more significant differences. The inferred radii ($R_2$) on the lighter component from the ST model are within the error bars on the general relativistic test, but the radius ($R_1$) of the heavier neutron star differs significantly. This is reflected in the comparison with the two NICER results from T. E. Riley et al. (2019) and M. C. Miller et al. (2019). It is interesting that while the GR result agrees with T. E. Riley et al. (2019), the $\beta = -6$ result has greater overlap with M. C. Miller et al. (2019). We conclude, therefore, that the differences between the two NICER analyses are within the biases introduced by alternate theories of gravity, and these biases are clearly greater for higher-mass neutron stars. We look at the Bayes factor to determine whether ST theory or GR is more strongly favored by the data. We find that $\mathcal{B}_{STT/GR} = 3.5 \pm 0.2$ ($\log_{10}\mathcal{B}_{STT/GR} = 0.54 \pm 0.03$). This demonstrates only a weak preference toward ST theory.

We turn now to the EOS and the underlying nuclear parameters. Both analyses select different EOSs within current measurement uncertainties, and hence might also provide different constraints on microphysical parameters in nuclear physics. In Figure 2, we show the posteriors for the most important NEPs discussed earlier: $K_{sat}$, $E_{sym}$, $L_{sym}$, and $K_{sym}$. Additionally, we also show the values of the speeds of sound at $3n_{sat}$, $c_s^2(3n_{sat})$. We find no observable difference between the inferred NEPs for observations with current detectors, and the posteriors for these parameters do not change appreciably between the different theories of gravity. Hence, we conclude that the different theories for gravity are not sensitive to the low-density EOS below $2n_{sat}$. This can be understood from the fact that the low-mass component in the merger, probing the low-density EOS, does not change significantly. Changes only appear at higher masses, and hence at higher densities.

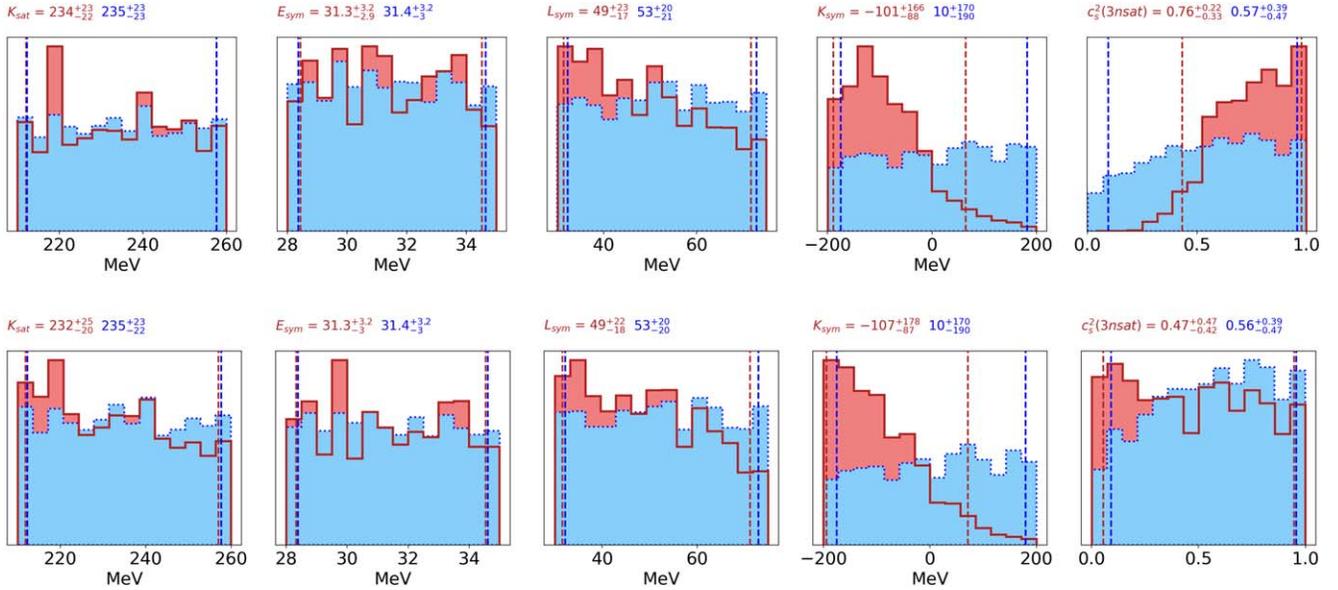

**Figure 2.** Posteriors on select nuclear properties ($K_{sat}$, $E_{sym}$, $L_{sym}$, $K_{sym}$, and $c_s^2$ at $3n_{sat}$) in GR (top) and ST theory (bottom). The prior is shown in blue and the posterior is shown in red. Dashed vertical lines show the 90th percentile credible interval.

However, the two different theories of gravity change our inference on parameters describing the high-density EOS. We find that the speed of sound at $3n_{sat}$ changes significantly, and the inferred high-density EOS is much softer for ST theory than for GR. The reason is that the EOS set generally extends to smaller radii in GR, which are ruled out by the inferred heavier neutron star. In contrast, in ST theory all EOSs in our set are curved toward larger radii, and almost all EOSs agree with the inferred second component. In this case, we observe no change from the prior, whereas there is a significant change in GR. For the speed of sound at $4n_{sat}$ and above, we do not find an observable difference between the two theories of gravity because these densities are above the central density of the heavier component, which on average is $3.2n_{sat}$.

Our results show that GR and theories with small deviations from GR may be indistinguishable using gravitational-wave analysis alone. They also show that deviations from GR can be absorbed into the nuclear EOS uncertainty, leading to biases in the nuclear EOS measurement. We have shown that measurements of the EOS around $3n_{sat}$ can be biased if potential deviations from GR are ignored. Thus, a measurement of the speed of sound at $3n_{sat}$ can potentially be useful for distinguishing between GR and ST theory. Furthermore, efforts are being made to understand if the domain of validity of nuclear effective field theories can be extended above $2n_{sat}$ (I. Tews et al. 2024). Investigating the EOS at these densities is therefore crucial in the upcoming years. Third-generation gravitational-wave detectors, planned to launch in the 2030s, are expected to improve both the number of events detected and the signal-to-noise ratio of nearby sources by at least an order of magnitude, making them key tools for precise measurements of the EOS (S. Hild et al. 2011; F. Acernese et al. 2014; M. Evans et al. 2021).

Additionally, our ability to probe the existence of phase transitions in dense matter will depend on our ability to measure the speed of sound (R. Somasundaram et al. 2023). From our results for the sound speed, we see that investigating modified theories of gravity is required for a comprehensive answer to such questions in the field of dense-matter physics.

As far as testing alternate theories of gravity is concerned, we have made several simplifications in our analysis. We have considered only the even-parity tensor perturbation, and neglected effects of scalar dipole radiation on the dynamics of the binary system, and hence on the GW signal (see, e.g., L. Bernard 2020). Since our aim here is to identify potential biases in EOS measurements, this suffices for our purposes. A detailed comparison of ST theories with standard GR will require these effects to be taken into account for different values of $\beta$ and $\varphi_\infty$.

### Acknowledgments

We thank Xisco Jiménez Forteza, Sayak Datta, Sumit Kumar, Pierre Mourier, and Gaston Creci for valuable discussions.

This work was supported by the research project grant "Fundamental physics from populations of compact object mergers" funded by VR under Dnr 2021-04195 and the research project grant "Gravity Meets Light" funded by the Knut and Alice Wallenberg Foundation under Dnr KAW 2019.0112. R.S. acknowledges support from the Nuclear Physics from Multi-Messenger Mergers (NP3M) Focused Research Hub, which is funded by the National Science Foundation under grant No. 21-16686, and from the Laboratory Directed Research and Development program of Los Alamos National Laboratory under project number 20220541ECR. I.T. was supported by the US Department of Energy, Office of Science, Office of Nuclear Physics, under contract No. DE-AC52-06NA25396, by the US Department of Energy, Office of Science, Office of Advanced Scientific Computing Research, Scientific Discovery through Advanced Computing (SciDAC) NUCLEI program, and by the Laboratory Directed Research and Development program of Los Alamos National Laboratory under project number 20230315ER. Our computations used the ATLAS computing cluster at AEI Hannover AEI (2017) funded by the Max Planck Society and the State of Niedersachsen, Germany. This research has made use of data or software obtained from the Gravitational Wave Open Science Center (https://gwosc.org/), a service of the LIGO

Scientific Collaboration, the Virgo Collaboration, and KAGRA. This material is based upon work supported by NSF's LIGO Laboratory, which is a major facility fully funded by the National Science Foundation, as well as the Science and Technology Facilities Council (STFC) of the United Kingdom, the Max-Planck-Society (MPS), and the State of Niedersachsen/Germany for support of the construction of Advanced LIGO and construction and operation of the GEO600 detector. Additional support for Advanced LIGO was provided by the Australian Research Council. Virgo is funded, through the European Gravitational Observatory (EGO), by the French Centre National de Recherche Scientifique (CNRS), the Italian Istituto Nazionale di Fisica Nucleare (INFN), and the Dutch Nikhef, with contributions by institutions from Belgium, Germany, Greece, Hungary, Ireland, Japan, Monaco, Poland, Portugal, and Spain. KAGRA is supported by the Ministry of Education, Culture, Sports, Science and Technology (MEXT), Japan Society for the Promotion of Science (JSPS) in Japan; National Research Foundation (NRF) and Ministry of Science and ICT (MSIT) in Korea; and Academia Sinica (AS) and National Science and Technology Council (NSTC) in Taiwan.

## ORCID iDs

Stephanie M. Brown  https://orcid.org/0000-0003-2111-048X
Badri Krishnan  https://orcid.org/0000-0003-3015-234X
Rahul Somasundaram  https://orcid.org/0000-0003-0427-3893
Ingo Tews  https://orcid.org/0000-0003-2656-6355